\documentclass[sigconf, review, authorversion]{acmart}

\usepackage[normalem]{ulem}

\AtBeginDocument{%
  \providecommand\BibTeX{{%
    \normalfont B\kern-0.5em{\scshape i\kern-0.25em b}\kern-0.8em\TeX}}}

\copyrightyear{2024} 
\acmYear{2024} 
\setcopyright{acmcopyright} 

\acmConference[WWW '24]{Proceedings of the ACM Web Conference 2024}{May 13--17, 2024}{Singapore, Singapore}

\acmISBN{978-1-4503-XXXX-X/18/06}

\usepackage{hyperref}
\usepackage{booktabs} 
\usepackage{color}

\begin{document}

\fancyhead{}
\title{PKG API: A Tool for Personal Knowledge Graph Management} 

\author{Nolwenn Bernard \qquad Ivica Kostric \qquad Weronika Łajewska \qquad \\ Krisztian Balog \qquad Petra Galu\v{s}\v{c}\'{a}kov\'{a} \qquad Vinay Setty \qquad Martin G. Skj{\ae}veland}
\affiliation{%
  \institution{University of Stavanger}
  \city{Stavanger}
  \country{Norway}
}
\email{{nolwenn.m.bernard, ivica.kostric, weronika.lajewska, krisztian.balog, petra.galuscakova, vinay.j.setty, martin.g.skjeveland}@uis.no}

\def\authors{}

\begin{abstract}

Personal knowledge graphs (PKGs) offer individuals a way to store and consolidate their fragmented personal data in a central place, improving service personalization while maintaining full user control. Despite their potential, practical PKG implementations with user-friendly interfaces remain scarce. This work addresses this gap by proposing a complete solution to represent, manage, and interface with PKGs. Our approach includes (1) a user-facing PKG Client, enabling end-users to administer their personal data easily via natural language statements, and (2) a service-oriented PKG API. To tackle the complexity of representing these statements within a PKG, we present an RDF-based PKG vocabulary that supports this, along with properties for access rights and provenance.

\end{abstract}



\maketitle

\section{Introduction}

A personal knowledge graph (PKG) is ``a resource of structured information about entities related to an individual, their attributes, and the relations between them''~\citep{Balog:2019:ICTIR}. A PKG offers the possibility to centrally store all information related to its owner such as personal relationship, preferences on food, and calendar data~\citep{Skjaeveland:2023:arXiv}.
This enables the delivery of highly personalized services while maintaining the owner's full control over their data. In today's digital world, where personal data is often fragmented across multiple accounts with different service providers, a PKG provides a solution for consolidating information.  Crucially, one of the  most essential features of a PKG is that the   individual is put in control of their data, allowing owners to determine what data is stored and what services have access to it~\citep{Skjaeveland:2023:arXiv}. 
Despite the clear potential of PKGs and the growing research interest around them~\citep{Balog:2022:SIGIRForum}, efforts have so far remained mostly on the conceptual level. Practical implementations, especially those that directly interface with users, are lacking. 
This paper aims to address that gap.

Similar to the concept of PKGs, Solid (Social Linked Data)~\citep{Sambra:2016:techrep} is an existing initiative that aims to put individuals in control of their own data. Solid allows users to store personal data in decentralized ``Pods'' (Personal Online Data Stores), giving them fine-grained control over which apps can access which portions of their data. However, Pods introduce a level of complexity that may pose challenges for ordinary web users.  Managing data within Pods requires a learning curve, and users accustomed to the simplicity of traditional services might find this transition difficult. Solid interfaces and applications have particularly been criticized for not being user-friendly, and compatibility issues between Pod providers and Solid apps lead to inconsistent user experiences.\footnote{See, e.g., discussions at  \url{https://www.reddit.com/r/solid/}}

In this work, we propose a user-friendly solution to managing PKGs, consisting of a web-based PKG Client and a service-oriented PKG API. 
To dramatically lower the barrier for end users, we let them administer and interact with their PKG via natural language statements, enabled by recent advances in Large Language Models (LLMs).
For example, a user might simply state a preference ``I dislike all movies with the actor Tom Cruise'' to be recorded in their PKG.
While this example is simplistic, we demonstrate that representing it in a PKG can actually become complex due to entanglements between different entities and relationships between them, such as \emph{all movies} and \emph{Tom Cruise}.
In order to tackle this challenge, we develop a PKG vocabulary on top of RDF to represent such statements both in natural language and as structured data. Furthermore, our vocabulary defines a set of properties, such as access rights and provenance, to enrich the statements.

In summary, the main contributions of this work are:
\begin{enumerate}
    \item A PKG vocabulary based on RDF reification, leveraging existing vocabularies, to represent statements in a PKG.
    \item A PKG API that implements both user-facing and service-oriented functionalities to access and manage a PKG. It includes a novel NL2PKG feature, enabling the translation of natural language statements to API calls.
    \item A web-based PKG Client to browse, expand, and manage a PKG, prioritizing simplicity, intuitive design, and visualization features for easy user understanding and control.
\end{enumerate}
Our complete solution along with a video demonstration may be found at \url{https://github.com/iai-group/pkg-api}.

\begin{figure*}[t]
    \centering
    \includegraphics[width=\textwidth]{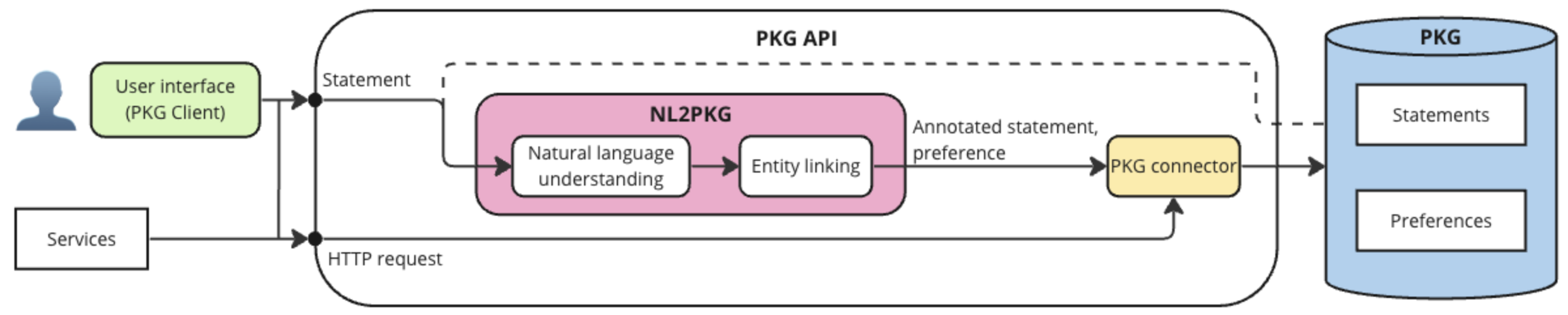}
    \caption{Overview of the PKG tooling developed in this work.}
    \label{fig:overview}
\end{figure*}

\section{Related work}
\label{sec:related}

While there are several definitions of PKGs~\citep{Balog:2019:ICTIR,Chakraborty:2022:WWW}, they all agree that PKGs can be seen as a specialized approach to a personal information management (PIM) system~\citep{Janssen:2022:JIR,Nasar:2011:STAIR}. PIM systems provide storage and access to personal data, with a focus on personal data control for third-party services. Several tools have been proposed, including  Solid~\citep{Sambra:2016:techrep}, MyData~\citep{Langford:2020:techrep}, and OpenPDS/SA~\citep{deMontjoye:2014:PLOSONE}, and are available as either free or commercial services. 
Compared with general PIM tools, a PKG stores data in the form of a knowledge graph (KG)~\citep{Hogan:2011:Surveys}, where information about entities and their relations is directly modeled as a graph structure. This structure, grounded in a pre-defined ontology, inherently supports operations such as summarization of personal information~\citep{Kang:2022:ICDE,Safavi:2019:ICDM} and cross-domain recommendation~\citep{Su:2023:arXiv}. 

Our proposed PKG API focuses on enabling  user-friendly interactions via natural language (NL). This requires a way to automatic translate NL statements to a structured query language that can directly interact with the PKG. Methods for such NL-to-structured-language translation traditionally focus on conversion to SQL~\citep{Yu:2018:EMNLP,Xu:2017:ArXiv}, with interactive approaches~\citep{Tian:2023:EMNLP} and LLMs~\citep{Jinyang:2023:AAAI,Scholak:2021:EMNLP} being the state of the art. A similar pattern is followed in the case of conversion from natural language to SPARQL, which can be used for querying knowledge graphs. While earlier methods were based, for example, on rules~\citep{Ngomo:2013:WWW} and machine translation~\citep{Yin:2021:FGCS}, recent studies start to explore LLMs~\citep{Taffa:2023:arXiv,Wang:2023:arXiv}. LLMs have also been utilized to translate NL queries to different API calls~\citep{Schick:2023:ArXiv,Patil:2023:ArXiv}.

\section{Overview and Architecture}

This work aims to provide a simple and user-friendly way to manage a PKG.
It comprises a web interface, i.e., PKG Client, and the PKG API.
The PKG API serves as a middleman between the PKG and both the PKG Client and external service providers.
It has two entry points: one for natural language statements and another one HTTP requests. 
See Fig.~\ref{fig:overview} for an overview.

A \emph{statement} is a fundamental unit of information in the PKG, containing at minimum its text content. For example, ``I dislike all movies with the actor Tom Cruise'' is a statement.
Statements can be further enriched with properties defined in the PKG vocabulary (Section~\ref{sec:datamodel}), following the Subject-Predicate-Object (SPO) model. In this example, we can extract ``I'' as the subject, ``dislike'' as the predicate, and ``all movies with the actor Tom Cruise'' as the object.
Note that our example is a particular type of statement, one that expresses a \emph{preference}. Preference statements are especially valuable for service providers seeking to personalize user experiences. Therefore, our PKG vocabulary (detailed in Section~\ref{sec:datamodel}) explicitly supports the representation of preferences, which are derived from statements via a derivation relationship.

\begin{figure}
    \centering
    \includegraphics[width=\textwidth, height=0.23\textheight, keepaspectratio]{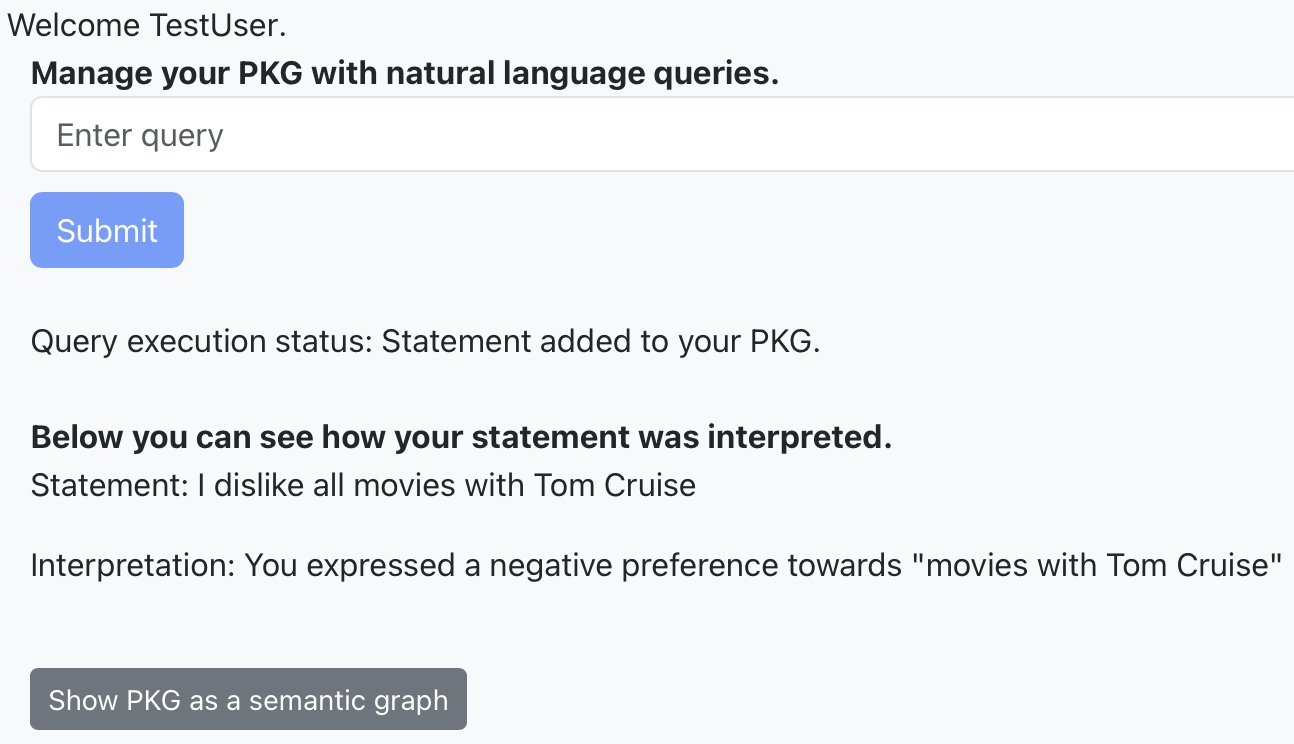}
    \caption{Screenshot of the home screen after submission of a natural language statement.}
    \label{fig:home_screen}
\end{figure}

The PKG Client is a web interface connecting users to their PKG. It is designed to be intuitive and user-friendly, aiming to make PKG administration accessible to a broad range of users.
The home screen features a form with a text area for users input \emph{natural language} statements, as well as an area to display the outcomes of these statements (Fig.~\ref{fig:home_screen}).
Additional screens within the interface provide forms for specific tasks like adding statements manually to the PKG and visualizing the PKG; these features are primarily designed for advanced users with knowledge of semantic web technologies.

When a natural language statement is received by the PKG API, whether from the PKG Client or external services, it is processed by the NL2PKG component (Section~\ref{sec:nl2api}), which performs two main steps.
First, a \emph{natural language understanding} step identifies the action to execute (e.g., adding a new statement), extracts properties (subject, predicate, object), and infers whether a preference is expressed (e.g., identifying a negative preference towards ``Tom Cruise''). At this stage, all the properties and preference are represented as text. 
Next, an \emph{entity linking} step attempts to resolve the properties and preference to IRIs (e.g., ``Tom Cruise'' to \texttt{<http://dbpedia.org/resource/Tom\_Cruise>}).
Note that these steps may be performed asynchronously on existing statements in case the natural language understanding and/or entity linking components are updated.
Once these steps are completed, the PKG Connector generates a SPARQL query with the annotated statement and preference, and sends it to the PKG.
Figure~\ref{fig:rdf-tom-cruise} illustrates these steps.

In the case where the user or service provider decides to interact with the PKG using HTTP requests, the PKG API directly triggers the PKG Connector to create and send the corresponding SPARQL query to the PKG. 

\section{PKG Vocabulary}
\label{sec:datamodel}

The PKG vocabulary is used for expressing all statements to be kept in a PKG.
It is specified as a set of SHACL shapes~\citep{Knublauch:2017:W3C}
over existing RDF vocabularies such as
RDF~\citep{Brickley:2014:W3C},
SKOS~\citep{Miles:2009:W3C},
PAV~\citep{Ciccarese:2014:web},
and The Weighted Interests Vocabulary~\cite{Brickley:2010:web},
including custom vocabulary terms necessary for expressing access rights to stored statements.

The main design idea behind the vocabulary is to provide a simple data model with which one can represent all kinds of incoming statements, and allow for incremental post-processing of statements to increase their quality and precision.
The core modeling pattern is standard RDF reification~\citep{Brickley:2014:W3C}: a statement is represented by an instance of \texttt{rdf:Statement} where the original statement in text is represented by a literal annotation on the statement. The extracted subject, predicate and object
are connected to the \texttt{rdf:Statement} using \texttt{rdf:subject}, \texttt{rdf:predicate}, and \texttt{rdf:object}, respectively, and can either be represented directly as an IRI, or, in the case that an appropriate IRI is not found, as an instance of \texttt{skos:Concept} with the extracted text as a literal annotation.

Instances of \texttt{rdf:Statement} and \texttt{skos:Concept} are further analyzed and can, if a match is found, be related to other known resources from the PKG or from external KGs using, e.g., the SKOS properties \texttt{skos:related}, \texttt{skos:broader}, or \texttt{skos:narrower}.
Also, the analysis may amend statements by asserting a preference the statement's subject has towards the object.
Additional semantic descriptions may be added to statements and concepts at the discretion and capabilities of the analysis tools, e.g., concepts like ``All movies with the actor Tom Cruise'' could be expressed as being a subclass of or equivalent to a, possibly complex, constructed OWL~\citep{W3COWLWorkingGroup:2012:W3C} class expression.
However, this is outside the scope of our current implementation.
Every \texttt{rdf:Statement} are assumed to be annotated with provenance information following the PAV Ontology.
Finally, the PKG vocabulary enables straightforward access control at the \texttt{rdf:Statement} level by explicitly stating which services have read and write access to the statement using the properties \texttt{pkg:readAccessRights} and \texttt{pkg:writeAccessRights}.

\begin{figure}
    \centering
    \includegraphics[width=\columnwidth]{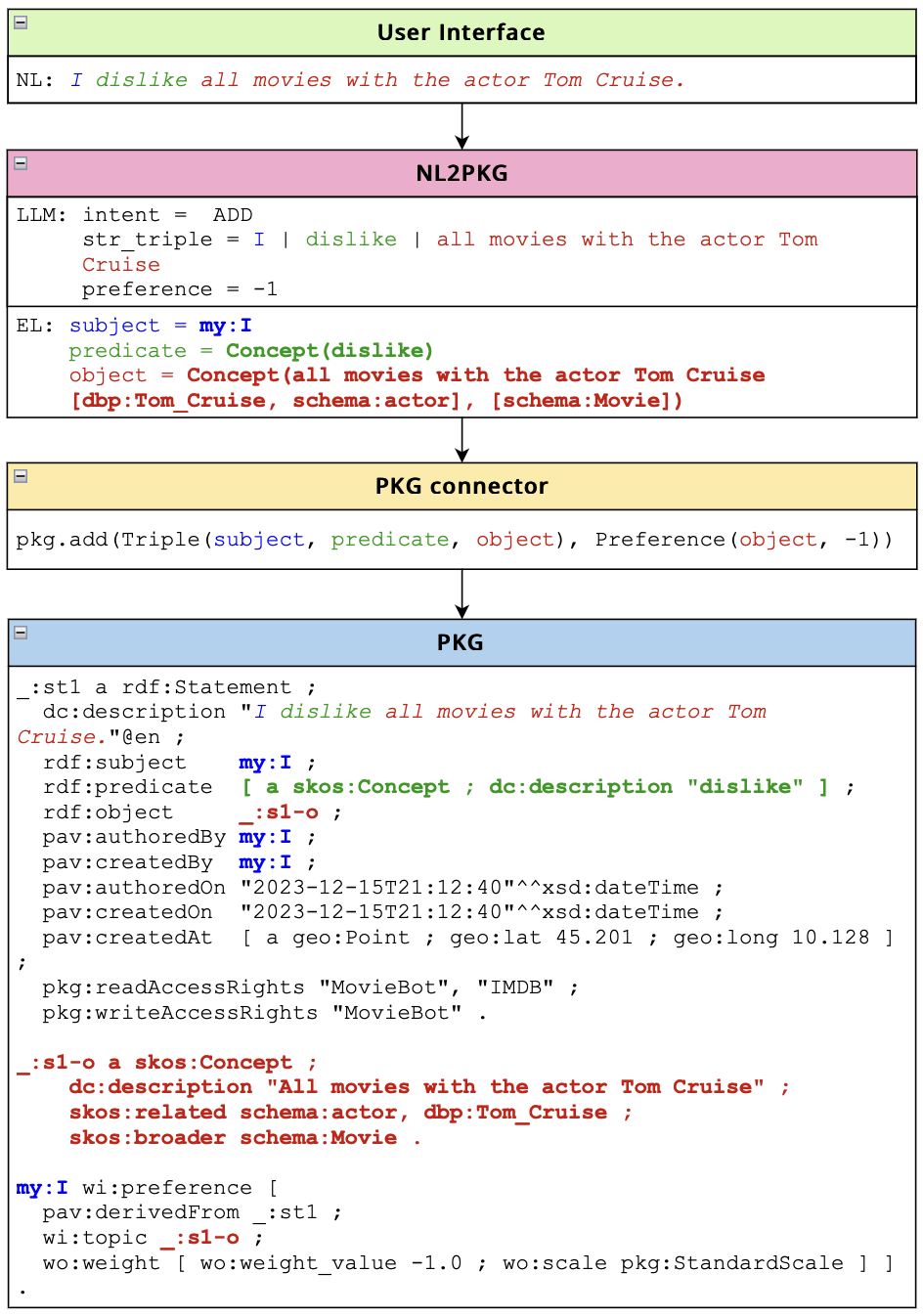}
    \caption{Life of a statement from NL to PKG.}
    \label{fig:rdf-tom-cruise}
\end{figure}

The bottom block of Fig.~\ref{fig:rdf-tom-cruise} demonstrates the use of the PKG vocabulary.
It is available at its namespace IRI: \url{http://w3id.org/pkg/}, including SHACL shape definitions, documentation, and examples.

\section{Natural Language to PKG}
\label{sec:nl2api}

To facilitate user-friendly interactions with PKGs, we present a two-stage NL2PKG approach that translates natural language statements to API calls that perform operations on the PKG, such as storing stated preferences or retrieving previous statements. 

In the first stage, we leverage LLMs to classify user intent, extract an SPO-triple, and identify whether a preference was expressed in the NL statement.
Intents specify the desired action on the PKG: (a) \texttt{ADD} inserts a statement, (b) \texttt{GET} retrieves matching statements, (c) \texttt{DELETE} removes a statement, and (d) \texttt{UNKNOWN} handles unrecognized statements. 
Preferences are represented as +1 (positive) or -1 (negative) and are relevant for
\texttt{ADD} intents in statements expressing user likes or dislikes. For example, ``Bob likes Oppenheimer'' translates to an \texttt{ADD} intent, inserting the triple $\langle Bob, Likes, Oppenheimer\rangle$ into the PKG with a preference of +1. 
Specifically, we employ pre-trained LLMs with few-shot chain-of-thought reasoning prompts. Separate prompts are used for intent classification, SPO-triple extraction, and preference identification.\footnote{The prompts used can be found here: \url{https://github.com/iai-group/pkg-api/tree/main/data/llm_prompts/cot}}

In the second stage, we employ an entity linker to resolve the SPO elements, which are initially extracted in their surface form representations. They are transformed into normalized entities and relations congruent with external KGs, such as DBpedia. Note that the subject element most commonly belong in the user's private circle (e.g., ``I'' and ``my mom''), thus, we argue that is should be resolved using an entity linker specific to the PKG.  

\section{Implementation}

Our solution contains two main components: (1) a PKG API served as a RESTful API with a backend server based on Flask, and (2) a user interface, PKG Client, implemented as a React application. 

Central to the PKG API's functionality is the NL2PKG module. 
In our demo, we use the Ollama framework\footnote{\url{https://ollama.ai}} to deploy and experiment with Llama2-7b and Mistral-7b as LLMs, with the latter being the default option based on a set of preliminary experiments.\footnote{We do not use proprietary models like GPT-4 or Gemini, due to privacy reasons.} 
For entity linking, we offer both REL~\citep{vanHulst:2020:REL}, as our default, and DBPedia Spotlight~\citep{Mendes:2011:ISemantics} as an alternative.
The code is designed to be modular to allow for easy experimentation with different LLM-based annotators and entity linkers in the future.

Natural language statements processed by the NL2PKG module are further handled by the PKG Connector responsible for the creation and execution of SPARQL queries against the PKG. The PKG Connector uses a dedicated Python package, RDFLib,\footnote{\url{https://rdflib.readthedocs.io/}} for generating and executing SPARQL queries in RDF format.

\section{Conclusion}
\label{sec:concl}

Personal knowledge graphs hold the potential to be useful tools for organizing and providing personal information. As the volume of digital data continues to grow, alongside the number of services that can utilize it, the need for user-centric management tools becomes ever more pressing. Recognizing that  existing tools are often too complex to be used by non-expert users, we focused on developing a robust internal data representation for PKGs, paired with an API and a user-friendly PKG Client. A key novelty of our approach is enabling users to interact with their PKG directly through natural language statements. Our open-source demo showcases the viability of this concept with a particular focus on understanding and representing user preferences. 
This work represents a major step forward in the practical realization of PKGs, opening avenues for research into both intuitive user-centric interaction methods and broader applications.

\begin{acks}
    This research was partially supported by the Norwegian Research Center for AI Innovation, NorwAI (Research Council of Norway, project number 309834).
\end{acks}

\bibliographystyle{ACM-Reference-Format}
\bibliography{www2024-pkg_api_demo.bib}

\end{document}